\documentclass[useAMS,usenatbib]{emulateapj}
\def\simless{{\th \rlap{\raise 0.5ex\hbox{$\scriptstyle  {<}$}}
    {\lower 0.3ex\hbox{$\scriptstyle  {\sim}$}} \th }}  %< or of order
\def\simgreat{{\th \rlap{\raise 0.5ex\hbox{$\scriptstyle  {>}$}}
    {\lower 0.3ex\hbox{$\scriptstyle  {\sim}$}} \th }}  %> or of order
\def\greateq{{\th \rlap{\raise 0.5ex\hbox{$\scriptstyle  {>}$}}
    {\lower 0.3ex\hbox{$\scriptstyle  {-}$}} \th }}  %< or of order
\def\lesseq{{\th \rlap{\raise 0.5ex\hbox{$\scriptstyle  {<}$}}
    {\lower 0.3ex\hbox{$\scriptstyle  {-}$}} \th }}  %> or of order
\def\th{\thinspace}
\def\ts{{\raise 0.3ex\hbox{$\scriptstyle {\th \sim \th }$}}}

\usepackage{epsfig}

\newcommand{\RXTE}{\textit{RXTE}}

\def\hide#1{}

\begin{document}

%\shortauthors{Patruno \etal}
\shorttitle{Spin evolution of IGR J00291}

%\submitted{Submitted to ApJ Letters}
\title{The Accreting Millisecond X-ray Pulsar IGR J00291+5934: Evidence for a Long Timescale Spin Evolution}
\author{Alessandro Patruno \altaffilmark{1}}
\altaffiltext{1}{Astronomical Institute ``Anton Pannekoek,'' University of
Amsterdam, Science Park 904, 1098 SJ Amsterdam, Netherlands}

\email{a.patruno@uva.nl}

\begin{abstract}

Accreting Millisecond X-ray Pulsars like IGR J00291+5934 are important
because it is possible to test theories of pulsar formation and
evolution.  They give also the possibility to constrain gravitational
wave emission theories and the equation of state of ultra dense
matter. Particularly crucial to our understanding is the measurement
of the long term spin evolution of the accreting neutron star.  An
open question is whether these accreting pulsars are spinning up
during an outburst and spinning down in quiescence as predicted by the
\textit{recycling scenario}. Until now it has been very difficult to
measure torques, due to the presence of fluctuations in the pulse
phases that compromise their measurements with standard coherent
timing techniques. By applying a new method, I am now able to measure
a spin up during an outburst and a spin down during quiescence. I
ascribe the spin up ($\dot{\nu}_{su}=5.1(3)\times
10^{-13}\rm\,Hz\,s^{-1}$) to accretion torques and the spin down
($\dot{\nu}_{sd}=-3.0(8)\times10^{-15}\rm\,Hz\,s^{-1}$) to magneto dipole torques, as
those observed in radio pulsars. Both values fit in the
recycling scenario and I infer the existence of a magnetic field for
the pulsar of $B\simeq2\times10^{8}$ G. No evidence for an enhanced
spin down due to gravitational wave emission is found. The accretion
torques are smaller than previously reported and there is strong
evidence for an ordered process that is present in all outbursts that
might be connected with a motion of the hot spot on the neutron star
surface.

\end{abstract}

\section{Introduction}

The \textit{recycling scenario} (\citealt{alp82}, \citealt{rad82})
provides an evolutionary link between the young slowly rotating
pulsars and the old fast millisecond pulsars. 
The evolutionary phase during which the pulsar is spun up happens when
a neutron star in a binary accretes gas stripped from the donor
companion.  The gas is channeled onto the magnetic poles producing
X-ray pulses modulated at the rotational frequency of the neutron
star. When the accreting pulsar starts to spin in the millisecond
range it is called an Accreting Millisecond X-ray Pulsar (AMXP).

To establish the presence of a spin up process it is crucial the
measurement of torques. According to accretion theory, the excess
angular momentum brought by the accreting gas is responsible for the
acceleration of the neutron star rotation.  However, if the angular
momentum of the gas is not sufficiently high, the neutron star can be
spun down during accretion (propeller regime, see the seminal works of
\citealt{ill75}, \citealt{gho79} and \citealt{ust06} for a recent
study).  The magnitude of the spin up/down is correlated with the
amount of accreted matter, and hence with the X-ray flux. In the past,
several attempts have been made to measure a spin up/down in several
AMXPs, and a wealth of measurements are now available for at least
eleven out of the thirteen known AMXPs (see \citealt{wij04},
\citealt{pou06} and \citealt{dis08}). To accomplish this, the X-ray
pulse phases are measured and a timing model is fitted to represent
the orbital motion, the spin frequency and its first time
derivative. In this model one makes the assumption that the rotational
parameters of the neutron star (spin frequency and its derivatives)
are coincident with the observed pulse frequency and time derivatives
(see for example \citealt{gal05} and \citealt{bur07}).

However, large deviations from this model are usually seen in the
timing residuals. These deviations, referred to as X-ray timing noise,
represent some unmodeled component in the pulse phases which has not
yet a conclusive explanation.  Recently, it has been shown that timing
noise in at least 6 AMXPs is strongly correlated with the X-ray flux
(\citealt{pat09a}). The major conclusion was that
it is the pulse phase rather than its second time derivative (i.e.,
the pulse frequency derivative) to be correlated with the X-ray flux,
contrary to what predicted by accretion theory. A similar
problem for accretion theory was seen in XTE J1807-204
\citep{pat09b}, in which the pulse frequency derivative has no
correlation with the X-ray flux whereas the pulse phases are strongly
correlated at all timescales with the X-ray flux. This is particularly
convincing given the presence of large fluctuations in the X-ray flux
at different timescales.

The first claim for a detection of an accretion torque in an AMXP was
made by \citet{fal05} for the AMXP IGR J00291+5934 (henceforth
referred to as IGR J00291) during the first outburst extensively
observed by RXTE/PCA and INTEGRAL in 2004. Differently from many other
AMXPs, the timing residuals of IGR J00291 look quite smooth, and the
presence of timing noise is not as dramatic as in other
AMXPs~\citep{pat09a}. Therefore at a first sight, it looks quite
obvious to look for the presence of a spin up which manifests as a
parabolic trend in the pulse phases and can be measured according to
standard coherent timing techniques.  However, also the 2004 X-ray
flux of IGR J00291 shows a smooth decay in time, and no sudden flux
variations are observed. Therefore if a correlation between the X-ray
flux and the pulse phases is present, it is difficult to disentangle 
it from a parabolic variation due to a true spin up, which is expected
to be uncorrelated with the X-ray flux variations.

\citet{pat09a} studied the 2004 outburst of IGR J00291 and claimed
that already during this outburst there is a possible correlation
between flux and pulse phases as those seen in other AMXPs. Indeed,
these authors questioned the presence of a torque \textit{as strong
as} that detected by \citet{fal05}, although they did not quantify the
magnitude of the expected torque in IGR J00291.

In 2008 the AMXP IGR J00291 went in outburst again \citep{cha08a}.
This outburst was quite anomalous with respect to the previous one
observed in 2004, since it showed first a faint outburst with a peak
flux of about half the value of the 2004, and then went into a very
low flux level phase for more than 30 days. During this low level
activity phase, the source was not detected by RXTE and was marginally
detected by XMM-Newton \citep{lew10}. After this period, a new high
level activity episode was recorded: the flux slowly rose for about 6
days, before decaying again and entering into quiescence on a
timescale of approximately one week from the outburst peak. The 2008
outburst has shown therefore strong flux variations that might be
correlated with the pulse phases.

The behavior of the 2008 outburst is also very attractive for the
purpose of testing accretion theory and pulsars evolution.  Thanks to
the long baseline of the observations, it is possible to follow the
evolution of the spin parameters in IGR J00291 on a timescale of 4
years. Only for two other AMXPs it has been possible to accomplish
this: SAX J1808.4-3658 (\citealt{har08}, \citealt{har09},
\citealt{pat09a}) and Swift J1756-2508 \citep{pat10}. The spin of
these sources is consistent with very weak or no accretion torques
during the outbursts, and with a spin down during quiescence that can
be interpreted as a magneto dipole torque like that operating in radio
millisecond pulsars.

The plan of the paper goes as follows. 
In Section~\ref{redu} I give a detailed summary of the
observations used and explain the methodology applied to analyze the pulse
phases and the X-ray flux.

In Section~\ref{sca} I perform a detailed standard coherent analysis
of the pulse phases of IGR J00291. The assumption made in this section
is that accretion theory provides a good description of the
behavior of pulse phases and consequently of the neutron star spin
parameters, even if timing noise is present.  This analysis is made to
verify whether the minimal assumptions made in accretion
theory are sufficient to provide a satisfactory explanation of the
pulse phase behavior. I discuss inconsistencies in the results
obtained with this methodology.

In Section~\ref{cca} I study the correlations between X-ray
flux and pulse phase variations. I propose an extended version of the
method first appeared in \citet{pat09a} to measure the spin frequency
and accretion torques under the assumption that X-ray flux variations
\textit{have an effect} on pulse phases. I call this method
Correlation Coherent Analysis, as opposed to the Standard Coherent
Analysis (see Sections~\ref{scam} and~\ref{ccam}).  With this method
I am able to remove the effects of the X-ray flux variations from
the pulse phases and measure ``cleaned'' spin parameters.

In Section~\ref{disc} I discuss the implication of the measurement of
the spin period of IGR J00291. I first discuss the behavior of the
pulsar spin during quiescence (Sec.~\ref{spin-down}) and then I
consider the detection of a spin-up during the 2004 outburst
(Sec.~\ref{sat}). A discussion on the origin of X-ray flux and pulse
phase correlations is discussed in Section~\ref{mot}, where I suggest
a possible connection of the pulse phase variations with a hot
spot moving on the neutron star surface.

The spin-up timescale of IGR J00291 and the consequences on the lack
of detected sub millisecond pulsars and the spin distribution of
accreting pulsars are discussed in Sections~\ref{equ}.

Finally, the implications of the observed spin evolution are
summarized in the framework of the \textit{recycling scenario}
(Section~\ref{recy}).

\section{Pulsations and X-ray Light Curve: Data Reduction}\label{redu}

I use all {{\it RXTE} PCA} public data for the 2004 and 2008 outbursts
of IGR J00291 (Table~\ref{obsid}). I refer to \citet{jah06} for PCA
characteristics and {{\it RXTE} PCA} absolute timing. I used all
available Event 125 $\mu$s and Good-Xenon data\footnote{Care has to be
taken when combining Events 125 $\mu$s and Good-Xenon data, since a
rigid shift of $2^{-12}$ s is present in the 2004 data due to a bug in
an early version of the FTOOL xenon2fits (Markwardt private
communication)}, rebinned to 1/8192 s and in the 5--37 absolute
channel that maximizes signal-to-noise-ratio (S/N). The absolute
channels correspond to $\approx 2.5-16$ keV.  The time of arrivals
are corrected to the solar system barycenter (TDB timescale) by using
the best available astrometric position reported in \citet{rup04}.
An optical position has also been reported by \citet{tor08}
which differs by 0.25 arcsec from the determination of \citet{rup04}.
If the optical position is used, then a shift in pulse frequency and
frequency derivative of $4\times10^{-8}$ Hz and $10^{-14}$ Hz/s is
expected (see for example Eq. A1 and A2 in \citealt{har08}).
These shifts are small enough to not affect the results reported in the
paper.

The light curve is folded in data chunks of different length, between
$\sim\,1000$ and $\sim\,3000$ seconds, keeping only those with
S/N$>$3--3.3$\sigma$, giving $<$1 false pulse detection per source.
The presence of V709 Cas in the field of view of IGR J00291
  has no effect on the determination of the pulse phases.  Indeed,
  using the harmonic decomposition~\citep{boy85}, only the pulse
  phases at the very frequency of IGR J00291 are measured. The pulse
  amplitudes are instead strongly affected by the presence of V709
  Cas, whereas the contamination cancels out when calculating the
  ratio between the amplitudes and the $1\sigma$ error. I detect only
a significant number of pulsations at the fundamental frequency
($\nu$) and then fitted the phases with a Keplerian orbit plus a
linear and possibly a parabolic term representing $\nu$ and
$\dot{\nu}$ (see Section~\ref{scam} and ~\ref{ccam} for more details).

I constructed the X-ray light curve using the counts in PCA Absolute
channels 5-37 ($\approx 2.5-16$ keV). The background contribution
(calculated with the FTOOL \emph{pcabackest}) is subtracted from the
total counts.

%########### TABLE ###########################
\begin{table}
\caption{{{\it RXTE}} observations for the 2004 and 2008 outburst}
%\centering
\scriptsize
\begin{tabular}{cccccc}
\hline
\hline
\RXTE & & & &\\
\hline
Year & Start & End & Outburst Length & Program IDs \\
& (MJD) & (MJD) & (days) &\\
2004 & 53355.85  & 53342.28  & 13.6  & {\tt 90052}- {\tt 90425}&\\          
2008 & 54696.77  & 54691.94  &  4.8  & {\tt 93013}- {\tt 93435}&\\
2008 & 54742.89  & 54730.51  & 12.38 & {\tt 94408}\\
\hline
\hline
\end{tabular}
\label{obsid}
\end{table}
%########### END TABLE ####################### 

\begin{figure*}[t]
  \begin{center}
    \rotatebox{0}{\includegraphics[width=1.0\textwidth]{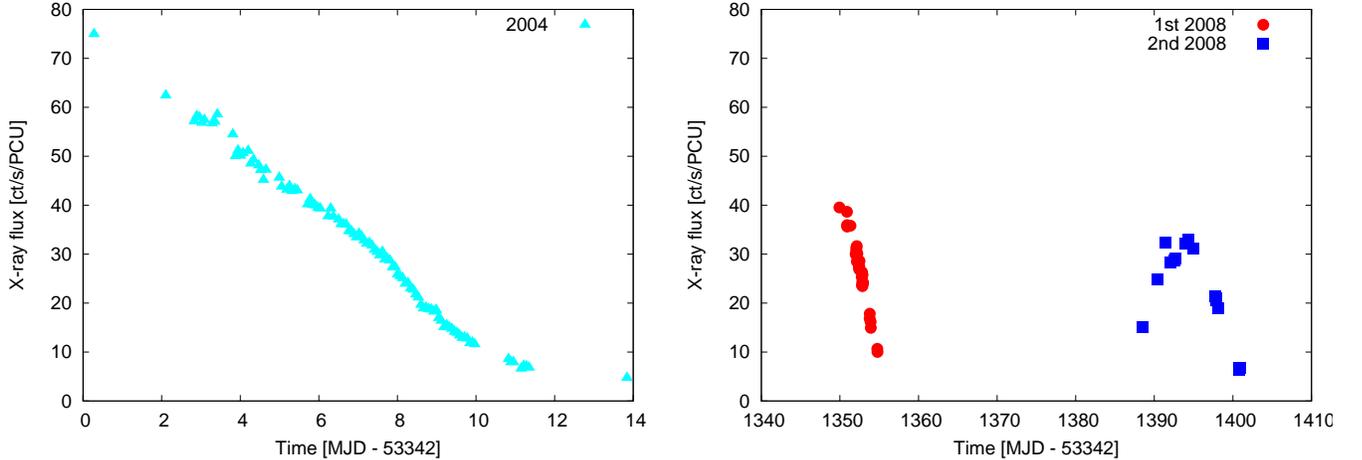}}
  \end{center}
  \caption{X-ray light curves of the two outbursts of IGR J00291.
The left panel shows the 2004 data, whereas the right one refers to the
2008 data. The y-scale is the same in the two plot, while the x-axis is not. 
In 2008 two faint outbursts are observed,with peak luminosities about half 
the value of 2004. The shape of 2004 and the first 2008 outburst is remarkably similar, 
whereas the second 2008 outburst shows a bell-shape light curve.  In the plot are 
shown only points in which pulsations are detected, which correspond to minimum counts
of approximately 8 ct/s/PCU. 
    \label{light curves}}
\end{figure*}

\subsection{X-ray light curves for the 2004 and 2008 outbursts}\label{lc}

The X-ray light curves for the 2004 and 2008 outbursts are shown in
Figure~\ref{light curves}.  The first outburst has an approximately
linear decay, with pulsations detected for 14 days, and a peak
luminosity of 75 ct/s/PCU in the 2.5-16 keV energy band, corresponding
to an unabsorbed flux of $9.5\times10^{-10}\rm\,erg/s/cm^{2}$
\citep{mar04} for an Hydrogen absorption column of
$N_{H}\sim7\times10^{21}\rm\,cm^{-2}$ and photon index
$\Gamma\simeq1.7$. \citet{fal05} used a comptonization model and
estimated a peak bolometric flux of
$\sim\,2.1\times10^{-9}\rm\,erg/s/cm^{2}$, corresponding to a
bolometric luminosity of $6.3\times10^{36}\rm\,erg\,s^{-1}$, assuming
a distance of 5 kpc (the distance is taken from \citealt{fal05}). In
this paper I fix the distance at 5 kpc in all calculations. The
corresponding mass accretion rate, given a neutron star mass of
$1.4M_{\odot}$ and an assumed efficiency of $\eta=0.15$, is therefore
$\dot{M}\sim 2\times10^{-9}M_{\odot}\,yr^{-1}$, according to the
expression $L_{bol}=\dot{M}\,\eta\,c^{2}$. The mass accretion rate
averaged over the entire 2004 outburst length is instead approximately
$\dot{M}_{outb}\simeq\,7\times10^{-10}M_{\odot}\,yr^{-1}$. Of course
uncertainties are present in this estimate, especially because no spectral
modeling has been done. Furthermore, there are uncertainties in the
conversion of X-ray to bolometric flux and on the assumed value of
$\eta$ for the conversion of rest mass energy into radiation, so these
values have to be taken with care and only as orders of magnitude
estimates. There is also a large uncertainty in the distance,
which has been proposed to be in the range 2-6 kpc by several authors
(\citealt{tor08},~\citealt{fal05},~\citealt{gal05}).

For the 2008 outburst, the analysis is separated for each of the two
weak outbursts observed.  Although the 2008 outburst can be considered
as a single episode of high level activity, I prefer to 
treat it as two separate outbursts since the coherent analysis will
 be also given separately (the reason of this choice is explained in
 Section~\ref{sca}).

The first weak outburst has a very similar shape as the 2004 outburst,
but a peak flux about half the 2004 value. The outburst appeared after 3.68 yr
from the 2004 outburst (Galloway et al. 2008).  Assuming that the
spectral shape remained the same as in 2004, the average mass
accretion rate is:
$\dot{M}_{outb}\simeq\,6\times10^{-10}M_{\odot}\,yr^{-1}$.  This
outburst lasted for approximately 5 days, after which a very low level
activity was recorded \citep{lew10}.

The second weak outburst appeared after 30 days and showed a different
shape than the 2004 and the first weak 2008 outburst.  It shows a slow
rise and a rapid decay that create interesting variations in the X-ray
flux shape that can be tested against the pulse phase variations.
Assuming that the spectral shape remained the same as in 2004, the
average mass accretion rate is:
$\dot{M}_{outb}\simeq\,5\times10^{-10}M_{\odot}\,yr^{-1}$.

Before the 2004 outburst, other two outbursts were detected in the
RXTE/ASM, with recurrence times of 2.80 and 3.23 yr (\citealt{rem04},
\citealt{gal08}).  The average recurrence time seems therefore to
slightly increase on the long timescale.  By using all these
information, I can calculate the long term average mass accretion
rate in IGR J00291 for two outburst/quiescence cycles:
\begin{eqnarray}
\left<\dot{M}\right>_{2004}&=&7\times10^{-10}\times\frac{13.6\,d}{365\,d}\frac{1}{3.20\,yr}\simeq8\times10^{-12}M_{\odot}\,yr^{-1}\nonumber\\
\left<\dot{M}\right>_{2008}&=&5\times10^{-10}\times\frac{12.4\,d}{365\,d}+6\times10^{-10}\nonumber\\&\times&\frac{4.8d}{365d}\frac{1}{3.20\,yr}\simeq7\times10^{-12}M_{\odot}\,yr^{-1}
\end{eqnarray}

Both values are in excellent agreement, and point towards an accretion
rate similar to that calculated for the AMXP SAX J1808.4-3658, which
has showed six outburst/quiescence cycles with an
$\left<\dot{M}\right>\simeq 10^{-11}M_{\odot}\,yr^{-1}$ \citep{bil01}.

\subsection{Old Method: Standard Coherent analysis}\label{scam}

Standard coherent methods (e.g.,~\citealt{tay92}) are based on folding
procedures and to $\chi^{2}$ minimization techniques with a model
describing the time evolution of the pulse phase $\phi\left(t\right)$
at the barycentric reference frame.  The pulse phases are then fitted
with a Keplerian orbit and a spin frequency and its first time
derivative (see for example \citealt{pat09b} and references therein
for a discussion of the method).
I performed this fit by using the standard coherent timing software
TEMPO2~\citep{hob06}.
The spin frequency derivative is then
associated with the accretion torque $N$ via the simple expression
given by accretion theory:

\begin{equation}
N=2\pi\,I\,\dot{\nu}_{su}
\end{equation}

If the orbital and astrometric components are correctly removed from
the pulse phases, one expects to see in the timing residuals a set of
independent values which are normally distributed around the zero
average with an amplitude that can be predicted by propagating the
Poisson uncertainties due to counting statistics.

I stress that this method uses the assumption that the pulse phases
vary in time only because of the rotational motion of the neutron star
around its spin axis and around the orbit, while no other effect is
taken into account. If some other process does affect the pulse
phases, then the Standard Coherent Analysis does not return realistic
physical parameters and their statistical uncertainties for the accreting
neutron star.

\subsection{New Method: Correlation Coherent Analysis}\label{ccam}

In this second method, one takes into account also possible \textit{additional}
effects others than the rotational motion of the neutron star on the
observed pulse phases. A first attempt to apply this method in AMXPs
was made by \citet{pat09a}. In that paper the authors considered the
possible influence of the X-ray flux on the pulse phase, suggesting
that mass accretion rate $\dot{M}$ induced hot spot motion dominates
the observed pulse phase variations.

The method I present here follows the same route, and searches for
higher/lower reference pulse frequencies than the one selected in
Standard Coherent Analysis by using the correlation between pulse
phases and X-ray flux. I make a non-trivial assumption on the nature
of the correlation: there is a \textit{linear} relation between the
pulse phases and the X-ray flux.  The selected pulse frequency is then
the one that minimizes the $\chi^{2}$ of the linear fit between phase
and flux, instead of the pulse frequency that minimizes the pulse
phase residuals in the fit with TEMPO2. The reason why I choose a
linear relation among all the possible choices is because this is the
simplest law, and it is by no means necessarily a universal law that
can be applied in all circumstances. One has to take this assumption
as the ``minimal hypothesis'', so that it is possible to verify
whether under the simplest circumstances it is possible to already obtain
results which are statistically better than Standard Coherent
Analysis.

Differently from \citet{pat09a}, that used a constant spin frequency
 model, I select the best pulse frequency \textit{and pulse frequency
 derivative} instead of varying simply the pulse frequency (see also
 \citealt{pat09b}). I scanned 3000 pulse frequency derivatives in the
 range $[-10^{-12}\rm\,Hz\,s^{-1},+10^{-12}\rm\,Hz\,s^{-1}]$ and 1000
 pulse frequencies for each outburst. Only for the 2004 outburst it is
 possible to detect a significant spin derivative, while in 2008 only
 non-constraining upper limits were calculated.  Since no significant
 pulse frequency derivative is detected in 2008, I refit the same data
 with a constant spin frequency model.

By applying this technique I obtain results for \textit{all the
outbursts} which show clear improvement in the $\chi^{2}$ of the fit
when using the Correlation Coherent Analysis rather than Standard
Coherent Analysis. It is relevant to note that when using the
Correlation Coherent Analysis instead of Standard Coherent Analysis,
the number of parameters used in the fit (and therefore the degrees of
freedom) are exactly the same for the two methods, so there is no risk
to over-fit the data when using the former method.

In the following Sections~\ref{sca} and \ref{cca} I discuss the
results and the implications of the two methods. If one applies
Standard Coherent Analysis, then the physical consequences seem
contradictory and require extraordinary explanations. If one considers
Correlation Coherent Analysis, then the results match with a high
degree of accuracy the predictions of Accretion Theory.
\begin{figure*}[t!]
  \begin{center}
    \rotatebox{-90}{\includegraphics[width=1.0\columnwidth]{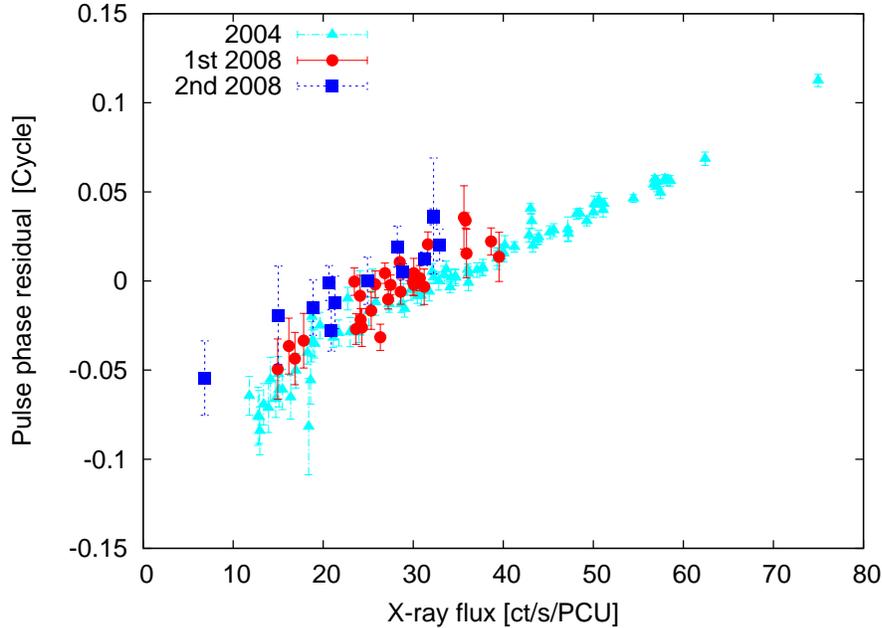}}
  \end{center}
  \caption{X-ray flux vs. pulse phase correlation for the 2004 and the two 2008 outbursts.
All three outbursts follow a correlation between flux and phase which is consistent with 
being the same in all three cases, within the statistical errors. 
The correlation has been fit with a linear relation between flux and pulse phase. 
The pulse phase shifts by almost 0.2 cycles during 2004 and by $\sim 0.1$ cycle during 2008, 
for both outburst episodes. The fit parameters and the $\chi^{2}$ of the fit are reported in Table~\ref{tabcorr}.
    \label{correlations}}
\end{figure*}
\section{Standard Coherent Analysis and Accretion Theory}\label{sca}

%I have applied Standard Coherent Analysis to the 2004 and 2008
%outburst data.  
The spin parameters for the three outbursts are shown
in Table~\ref{timsol}. The orbital parameters are instead shown in
Table~\ref{taborbit:2004},~\ref{taborbit:2008_1st}
and~\ref{taborbit:2008_2nd}.

According to this method of analysis the spin frequency between the
first and the second 2008 outburst decreases by $\sim0.7\rm\,\mu\,Hz$.
This means that during the first 2008 outburst, or in between the two
2008 outbursts, a strong spin down must have occurred.  However, a
spin down during the outburst is not a realistic hypothesis because we
detect positive spin frequency derivatives, firmly excluding any spin
down. According to this method of analysis, the required spin down
must have occurred in between the two 2008 outbursts and it needs to
be of the order of $-3\times10^{-13}\rm\,Hz\,s^{-1}$. This value
appears to be too large to be explained with a propeller scenario.
Indeed, at MJD 54703, during the 30 days of low level activity, an
XMM-\textit{Newton} observation was performed, with upper limits on
the 0.5-10 keV X-ray flux of $\sim 10^{-14} \rm\,erg\,s^{-1}\,cm^{-1}$
\citep{lew10}, which are almost 5 orders of magnitude lower than the
peak luminosity of the first 2008 weak outburst. This means that the
accretion level must have been extremely low, around
$10^{-14}\,M_{\odot}\,yr^{-1}$ or less, and this is not compatible
with the required mass that need to be ejected from the system (which
is of the order of a few $10^{-10}\,M_{\odot}\,yr^{-1}$) to explain
the $\sim0.7\rm\,\mu\,Hz$ spin frequency shift. Of course one can argue
that at the onset of the propeller a large amount of mass is still
present, but the X-ray luminosity is suppressed by the lack of
accretion on the neutron star surface. However, also the production of
X-rays in the inner accretion disk must be suppressed, since the X-ray
luminosity observed is comparable with the quiescent luminosity
\citep{cam08, tor08}.

Also a magnetic dipole induced spin down can be considered a quite
unlikely possibility since the spin down between the two 2008 weak
outbursts requires a magnetic field of at least $2\times10^{9}$ G and
there is no reason why such large spin down is not observed between
the 2004 and the first 2008 outbursts. Indeed, the spin
frequency at the end of the 2004 outburst is $598.89213111(5)$ Hz and
with a spin-down of $-3\times10^{-13}\rm\,Hz\,s^{-1}$ constantly
operating in the $\sim1336$ days of quiescence, the spin frequency
expected at the beginning of the first 2008 outburst should have been
$\sim598.8920965$ Hz. This is $34\mu\,Hz$ off the measured spin
frequency at the beginning of the first 2008 outburst, and \textit{hundreds}
of sigma away.

A final possibility to explain these mismatching spin frequencies is
via glitches occurring \textit{during quiescence}, since no signature
of glitches is seen in the timing analysis of the outbursts.
Glitches in AMXPs have never been observed and those detected in
millisecond (radio) pulsars are a rare phenomenon. In these latter
systems, the magnitude on the spin frequency variation during a glitch is
orders of magnitude smaller than in normal young pulsars
\citep{cog04}. There is only one detection of a glitch in a slowly
rotating accreting pulsars for the source KS
1947+300~\citep{gal04}. However, in this accreting pulsar the glitch
resulted in an \textit{acceleration} of the spin rotation, while we
need here a deceleration.  Therefore,
although a glitch cannot be completely ruled out, it appears an
unlikely explanation for the measured mismatching frequencies.

I ascribe this behavior of the spin frequencies to the presence of
timing noise as already suggested in \citet{pat09a}.  A further
consequence of this is that phase connection between the two 2008
outbursts is not a correct procedure of analysis. Indeed, although
certainly possible, the phase connection with the standard coherent
technique ignores the presence of timing noise and assumes that the
pulse phases are well behaved. 
%A phase-connection is therefore highly
%risky, and should not be used any further, unless one is completely
%sure that no timing noise is present in the data. 
Therefore the spin frequency that one
finds by phase connecting the two 2008 outburst in this way is just an
average value of the spin frequencies plus torques and contamination
of timing noise and as such it is a meaningless quantity. However,
this average value does not necessarily deviate from the true value by
a large amount. This depends on the weighted effect of several
factors, like the outburst length, the strength of timing noise and
the magnitude of torques during and in between outbursts, that might
also partially compensate each other and mislead the judgment on the
validity of the results. 

Another problem is related with the magnitude of the spin frequency
derivative detected in 2004. If one does not take into account the
effect of timing noise, then also this quantity, and hence the
inferred accretion torques, will appear higher than they really are.
I give here an example of this effect on the measured spin frequency
derivatives of IGR J00291. The 2004 outburst of IGR J00291 started on
December 3, and stopped on December 21.  \citet{gal05} analyzed the
data between December 3 to 6 and did not detect any spin frequency
derivative, with large upper limits due to the short baseline of the
observation.  \citet{fal05} and \citet{bur07} analyzed the pulsations
from December 7 to 21 and they both detect a consistent spin frequency
derivative between 8 and 12$\times10^{-13}\rm\,Hz\,s^{-1}$.  However,
if one splits the data differently, for example one analyzes the data
between December 3 and 10 and between 10 and the end of the outburst,
one finds $\dot{\nu}_{su}=3.6(5)\times10^{-13}\rm\,Hz\,s^{-1}$ and
$\dot{\nu}_{su}=24(5)\times10^{-13}\rm\,Hz\,s^{-1}$. So if one believes
accretion theory, the torque has \textit{increased} when the
flux was lower, completely contradicting the expectation that a
smaller amount of mass should bring less angular momentum and produce
a smaller and not a higher spin frequency derivative.  There might be
possible explanations for that, i.e., considering modifications to the
accretion theory, but taking into account the presence of timing noise
might be a more straightforward explanation. Indeed something similar
was observed in XTE J1807-294 \citep{pat09b} where the short-term spin
frequency derivatives were behaving in a way not predicted by accretion 
theory. The conclusion was that the observed spin frequency
derivatives were affected by red timing noise, so they were not
true spin frequency derivatives. The same conclusion
applies here. In the next Section I propose a method that takes into
account the effect of timing noise and calculates unbiased spin
frequencies and time derivatives.

\section{Correlation Coherent Analysis and Accretion Theory}\label{cca}

A possibility to explain the pulse frequency discrepancy of the
previous Section is via the presence of timing noise in the pulse
phases.  To take into account this, I have applied the Correlation
Coherent Analysis (see Section~\ref{ccam}) to the 2004 and 2008
outburst data.

I used the orbital solution as found in Section~\ref{sca}, since the orbit
is only marginally affected by the pulse phase noise, which operates
on completely different timescales than the orbital modulation.
Indeed the orbital period of IGR J00291 is $\sim\,2.4$ hr, while the
flux changes on timescales of days.  The very weak dependence of the
orbital parameters on timing noise was already noticed by
\citet{har08} and \citet{pat09b}, and is not relevant for compact binaries
in which the timing noise operates on timescales longer than the
orbital period. I refer to the aforementioned papers for a detailed
discussion of the problem and further references.

I fit a linear relation between the pulse phases $\phi$ and the X-ray
flux $f_{X}$: $\phi=A+B\cdot\,f_{X}$. The coefficients A and B are
reported in Table~\ref{tabcorr}, along with the spin frequency and its
first derivative found minimizing the $\chi^{2}$ of the
linear fit. All coefficients are consistent with being the same within
the statistical uncertainties. All the statistical errors are
calculated for a $\Delta\chi^{2}=1$. The correlations for the three
outbursts are shown in Figure~\ref{correlations}, where it is evident
that all three outbursts behave in a way which is consistent with
being the same. Of particular relevance is the fact that also the
second 2008 outburst follows a similar correlation as the 2004 and the
first 2008 outburst, even though the shape of this X-ray light curve is
very different from the other two. The pulse phases suggest 
a drift of approximately 0.2 cycles in 2004 and 0.1 cycles in the two 
2008 outbursts. 

There is still some unmodeled component in the fits, which is
particularly evident in 2004, where the pulse phases slightly deviate
towards the end of the outburst and affect the goodness of the fit.
However, the $\chi^{2}$ of the fits obtained with this method are
statistically much better than those obtained with standard coherent
analysis (see last column of Table~\ref{tabcorr} and \ref{timsol}).
The uncertainties found with this method are larger than those found
with standard coherent analysis because here I take into account also
the effect of timing noise in the fit. The statistical uncertainties
found in this way can be considered a good approximation of the true
uncertainties, differently from those obtained with standard coherent
timing techniques.

In a recent work, \citet{ibragimov} demonstrated that pulse
phase variations have an energy dependence in the AMXP SAX
J1808.4-3658.  Something similar might also happen in IGR J00291, so
it is useful to test whether the linear correlation and the best spin
frequency and frequency derivative have an energy dependence.  I split
the data in two energy bands, a soft band (2-7 keV) and an hard band
(8-16 keV) and repeated the entire procedure outlined above for the
full band (2-16 keV). The best spin parameters as well as the
coefficients A and B are consistent with being the same for the soft,
hard and full bands. However, since the photon statistics is degraded
when splitting the data in sub-bands, the statistical errors in the
soft and hard bands are much larger than in the full band, so that a
possible energy dependence cannot be firmly excluded.

\begin{figure*}[t]
  \begin{center}
    \rotatebox{0}{\includegraphics[width=0.6\textwidth]{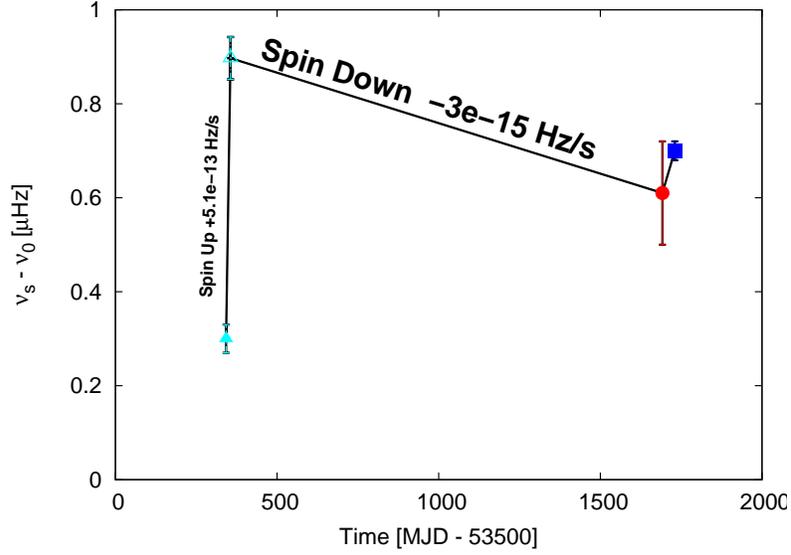}}
  \end{center}
  \caption{Long term spin frequency evolution of the pulsar in IGR
J00291.  The spin frequencies are plotted with an offset
$\nu_{0}=598.8921300$ Hz.  The symbols and colors are the same as
those in Figure~\ref{light curves} and \ref{correlations}. The red
circle and blue square are the average spin frequencies as measured in
the first and second 2008 outburst. The cyan triangle is the spin
frequency at the beginning of the 2004 outburst and the open cyan
triangle is the spin frequency at the end of the 2004 outburst, after
a spin up has taken place.  The spin frequency during the first 2008
weak outburst has decreased with respect to 2004,
and requires a spin down in quiescence of $-3\times10^{-15}\rm\,Hz\,s^{-1}$.
    \label{spin-evolution}}
\end{figure*}

The long term spin frequency evolution is reported in
Figure~\ref{spin-evolution}.  During the 2004 outburst there is a
detection of a spin up, while in the two weak 2008
outbursts it is only possible to set confidence intervals for the
non-detections which are consistent with being the same within the
statistical uncertainties.  The long term spin frequency evolution
requires a constant spin down during quiescence.  A detailed
discussion follows in the next Section.

%\centering
%\scriptsize
%########### TABLE ###########################
\begin{table*}
\caption{STANDARD COHERENT ANALYSIS: SPIN PARAMETERS FOR IGR
J00291+5934}

%\centering
\scriptsize
\begin{center}
\begin{tabular}{ccccc}
\hline
\hline
Outburst &  Spin Frequency & Spin Frequency Derivative & Epoch & $\chi^{2}/dof$\\
         &   [Hz]          &  [$10^{-13}\rm\,Hz\,s^{-1}$]& [MJD]  &\\
\hline
2004     &  598.89213045(1) & 5.6(3) & 53342.27  & 309.15/88 \\
1st 2008 &   598.89213061(8) & 12.3(4) &  54692.00 & 46.39/25 \\
2nd 2008 &   598.89213046(5) & 5.7(8) & 54730.50 &  26.55/10\\
\hline
\end{tabular}
\end{center}
\label{timsol}
\end{table*}
%########### END TABLE #######################

\begin{table}
  \caption{2004 Orbital parameters for IGR J00291+5934 }

\begin{center}
\begin{tabular}{lll}
\hline
\hline
\hline
Orbital Period [s]& 8844.080(2) \\
Projected Semi-major Axis [lt-ms] &  64.995(2)\\
Time Of Ascending Node [MJD] & 53345.1619277(5)\\
\hline
\hline
\end{tabular}
\end{center}
\label{taborbit:2004}
\end{table}

%\scriptsize
\begin{table}
  \caption{First 2008 outburst: Orbital Parameters for IGR J00291+5934 }

\begin{center}
\begin{tabular}{lll}
\hline
\hline
\hline
Orbital Period [s]& 8844.069(12) \\
Projected Semi-major Axis [lt-ms] &  64.987(4)\\
Time Of Ascending Node [MJD] & 54692.041113(3)\\
\hline
\hline
\end{tabular}
\end{center}
\label{taborbit:2008_1st}
\end{table}

%\scriptsize
\begin{table}
  \caption{Second 2008 outburst: Orbital Parameters for IGR J00291+5934 }

\begin{center}
\begin{tabular}{lll}
\hline
\hline
\hline
Orbital Period [s]& 8844.075(6) \\
Projected Semi-major Axis [lt-ms] &  64.993(5)\\
Time Of Ascending Node [MJD] & 54730.529224(4)\\
\hline
\hline
\end{tabular}
\end{center}
\label{taborbit:2008_2nd}
\end{table}

%########### TABLE ###########################
\begin{table*}
\caption{CORRELATION COHERENT ANALYSIS: SPIN PARAMETERS FOR IGR J00291+5934}

%\centering
\scriptsize
\begin{center}
\begin{tabular}{ccccccccc}
\hline
\hline
Outburst & A & $\sigma_{A}$ & B & $\sigma_{B}$ & Spin Frequency & Spin Frequency Derivative [$10^{-13}\rm\,Hz\,s^{-1}$]& Epoch & $\chi^{2}/dof$\\
\hline
2004     & -0.08 & 0.01 & 0.0024 & 0.0003 & 598.89213030(3) &  5.1(3) & 53342.27  & 184.23/88 \\
1st 2008 & -0.10 & 0.04 & 0.0034 & 0.0013 & 598.89213061(11) & $[0;18]$ (95\% c.l.) & 54692.00 & 46/26\\
2nd 2008 & -0.08 & 0.01 & 0.0032 & 0.0004 & 598.89213070(2) & $[1;11]$  (95\% c.l.)& 54730.50 &  19.4/11\\
\hline
\end{tabular}
\end{center}
\label{tabcorr}
\end{table*}
%########### END TABLE #######################

\section{LONG TERM SPIN EVOLUTION}\label{disc}

\subsection{Pulsar spin down in quiescence}\label{spin-down}

The spin-down evolution of the neutron star in IGR J00291 is reported
in Figure~\ref{spin-evolution}.  The pulsar spin frequency requires a
spin down between the end of the 2004 outburst and the beginning of
the first 2008 outburst.  
Since it is not possible to find a significant torque in the first and
second 2008 outbursts, the reported spin frequencies refer to a
constant spin frequency model. The value of the spin down is
$\dot{\nu}_{sd}\simeq-(3\pm0.8)\times10^{-15}\rm\,Hz\,s^{-1}$.

It is not possible to verify whether a spin down is also required
between the two 2008 weak outbursts, since the uncertainty on the spin
frequencies of the first 2008 outburst is too large ($\simeq
10^{-7}\rm\,Hz$). 

While other effects might also contribute to the spin down, the
rotating neutron star magnetic field is always present and causes a
continuous emission of low frequency radiation. By using the
force-free MHD approximation of \citet{spi06}, the determination of
$\dot{\nu}_{sd}$ provides an upper limit on the dipole moment:
\begin{eqnarray}
  \mu & < & 10^{26} 
    \left(1 + \sin^2\alpha\right)^{-1/2}
    \nonumber\\ & & \times
    \left(\frac{I}{10^{45}\ {\rm g\ cm^2}}\right)^{1/2}
    \left(\frac{\nu}{600\ {\rm Hz}}\right)^{-3/2}
    \nonumber\\ & & \times
    \left(\frac{-\dot{\nu}_{sd}}{3\times10^{-15}\ {\rm Hz\ s^{-1}}}\right)^{1/2}
    \textrm{ G cm}^3\, .
  \label{spindowndipole}
\end{eqnarray}

For the extreme values of $\alpha=90^{\circ},0^{\circ}$, and considering
all sources of uncertainty (colatitude, period and
period derivative),  the maximum
dipole magnetic field at the poles is
\begin{equation}
B_{sd}=[1.5; 2.0]\pm0.3\times10^{8}\rm\,G
\end{equation}

\begin{figure}[t]
  \begin{center}
    \rotatebox{0}{\includegraphics[width=1.0\columnwidth]{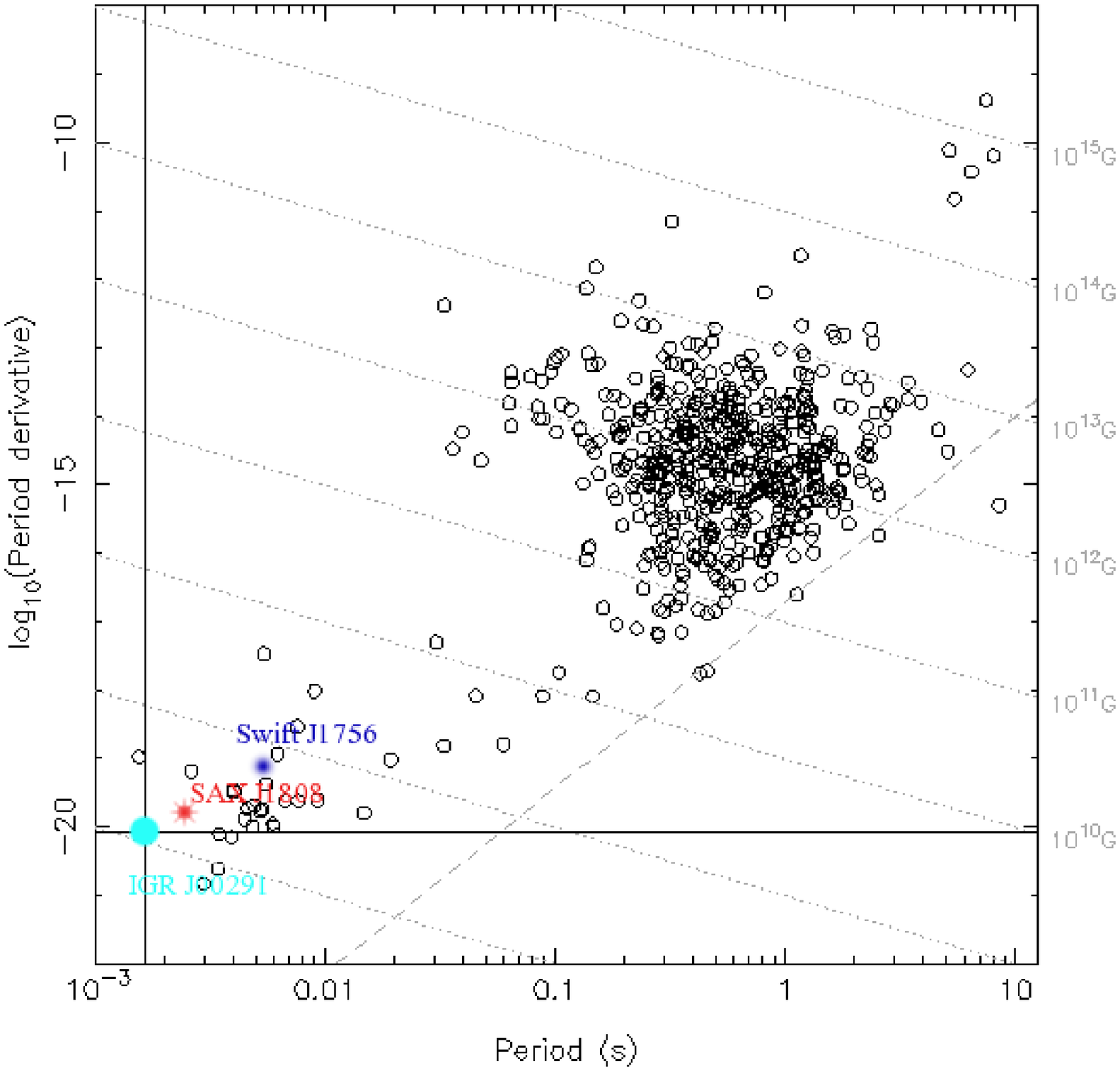}}
  \end{center}
  \caption{P-$\dot{P}$ diagram for radio pulsars. The cross represents
 the position of the pulsar in IGR J00291 with period and period
 derivative as determined in Section~\ref{spin-down} under the
 assumption that the observed spin down is caused by magneto dipole
 torques.  I plot also the position of SAX J1808.4-3658 (red star,
 \citealt{pat09a}) and Swift J1756.9-2508 (blue circle,
 \citealt{pat10}). The latter has no detected spin down during
 quiescence, and therefore the position is determined by using an
 upper limit on the spin down. 
    \label{p-pdot}}
\end{figure}

If the pulsar in IGR J00291 switched on as a radio pulsar during quiescence, 
then its position on the $P-\dot{P}$ would fall exactly in the expected
region occupied by radio millisecond pulsars (see Figure~\ref{p-pdot}).

\subsubsection{The Pulsar Magnetic Field}\label{mag}

Beside IGR J00291, there are only two other AMXPs in which it was
possible to constrain the magnetic field from the long term spin
frequency behavior: SAX J1808.4-3658 (\citealt{har08} and
\citealt{har09}) and Swift J1756.9-2508 \citep{pat10}.  For SAX J1808
a spin down of $-5.5\times10^{-16}\rm\,Hz\,s^{-1}$ was found.  For
Swift J1756 only upper limits on the spin down were given. 

The strength of the spin down in quiescence and hence of the magnetic
field of SAX J1808 was also revised by \citet{pat09a}, who applied a
similar technique to that reported in this paper, and found a spin
down of $-2\times10^{-15}\rm\,Hz\,s^{-1}$ and a magnetic filed of
$\simeq2-2.8\times 10^{8}$G, slightly larger than the $1.4\times
10^{8}$G reported in \citet{har09}.

The magnetic fields $B$ of the three pulsars are therefore:
\begin{itemize}
\item IGR J00291+5934: $[1.5;2.0]\pm0.3\times10^{8}$G\\
\item SAX J1808.4-3658: $[2.0;2.8]\pm0.2\times10^{8}$G\\
\item SWIFT J1756.9-2508: $[0.4;9]\times10^{8}$G (95\% c.l.)\\
\end{itemize}

The ranges $[1.5;2.0]$ and $[2.0;2.8]$ reflect the indetermination of
the colatitude of the magnetic pole, while the errors on each
determination is calculated by propagating the errors on the spin
parameters in Eq.~(\ref{spindowndipole}). The $B$ field of IGR J00291,
SAX J1808, and Swift J1756 is perfectly compatible with the minimal
hypothesis that only magneto dipole torques are at work. Under this
assumption, the derived magnetic fields are exact values and not upper
limits. There is no clear evidence for an alternative/additional
mechanism to explain the observed spin down other than magneto dipole
torques.

\subsection{Pulsar spin up during outbursts}\label{sat}

According to accretion theory, a pulsar accreting from a disk
will experience a positive torque (spin-up) when the magnetospheric radius
$r_{m}$ is smaller than the corotation radius $r_{co}$. The latter is
defined as the position at which the gas in the accretion disk has the
same angular velocity of the neutron star:
\begin{equation}
r_{co}\simeq 17\rm\,km\times\left(\frac{P_{spin}}{1 ms}\right)^{2/3}\left(\frac{M}{1.4\,M_{\odot}}\right)^{1/3}\label{rco}
\end{equation}  
For a pulsar spinning at $\sim600$ Hz like IGR J00291, the corotation
radius is at $\sim\,24$ km, which is within approximately one stellar
radii from the neutron star surface.

The magnetospheric radius is related to the neutron star magnetic
dipole moment $\mu$, to the neutron star mass $M$ and to the mass
accretion rate $\dot{M}$ via the equation:
\begin{eqnarray}
r_{m}&\simeq&35\rm\,km\,\xi\left(\frac{\mu}{10^{26}\rm\,G\,cm^{3}}\right)^{4/7} \nonumber\\
    &\times&\left(\frac{10^{-10}\,M_{\odot}\,yr^{-1}}{\dot{M}}\right)^{2/7}\left(\frac{1.4\,M_{\odot}}{M}\right)^{1/7}\label{rm}
\end{eqnarray}

The accretion disk model-dependent factor $\xi$ lies in the range
$\sim\,0.1-1$ (see e.g., \citealt{psa99}, \citealt{gho79} and
\citealt{bil97}). By assuming $M=1.4\,M_{\odot}$ and using the 2004
average mass accretion rate as calculated in Section~\ref{lc} I obtain
$r_{m}=2-20\rm\,km$, for $0.1<\xi<1$ . Values of
$r_{m}$ smaller than 8-10 km are non-physical because of the presence
of the hard surface of the neutron star. Nonetheless, given the large
uncertainties in the determination of $\dot{M}$ (distance, accretion
efficiency and bolometric flux) it is over-simplistic to favor values
of $\xi$ closer to one and the calculations have to be interpreted
just as order of magnitude estimates of the quantities involved.

The condition $r_{m}=r_{co}$ to enter the propeller
regime\footnote{More recent works (like \citealt{rap04} and
  \citealt{spr93}) show that the onset of the propeller is more
  complicated than simply $r_{m}=r_{co}$. However, this does not
  substantially affect the argument used in this paper.} is met when
$\dot{M}\simeq 10^{-10}M_{\odot}\,yr^{-1}$, which corresponds to an
RXTE count rate of $\sim5\rm\,ct/s/PCU$, in excellent agreement with
the minimum count rate at which significant pulsations are detected:
$\sim 8\rm\,ct/s/PCU$.  Although this value depends on several
additional sources of uncertainty , it shows that the magnetospheric
radius really lies within $r_{co}$ during the whole period in which
pulsations are observed from 2004 to 2008.

The maximum observed X-ray flux, and hence the maximum mass accretion
rate as determined in Section~\ref{lc} is
$\dot{M}=2\times10^{-9}M_{\odot}\,yr^{-1}$.  The magnetospheric radius
at this flux is $r_{m}\simeq 1.5-15\rm\,km$, for $0.1<\xi<1$. 

To check the consistency of our results I can compare the
maximum expected dipole magnetic filed calculated with accretion theory
with the magneto dipole torque inferred independently from the spin down
(Section~\ref{spin-down} and \ref{mag}). 
Under the minimal assumption that $r_{m}=r_{co}$:

\begin{eqnarray}
\mu_{max}&=&2.0\times 10^{23}\rm\,G\,cm^{-3}\left(\frac{r_{co}}{\xi}\right)^{7/4} \nonumber\\
         &\times&\left(\frac{\dot{M}}{10^{-10}\,M_{\odot}\,yr^{-1}}\right)^{1/2}\left(\frac{1.4\,M_{\odot}}{M}\right)^{1/4}
\end{eqnarray}

The value of $\mu_{max}$, and hence of the magnetic dipole field
depends on the choice of $\xi$ and on the average outburst
mass accretion rate. By choosing the least constraining values for
$\xi=0.1$ and
$\left<\dot{M}\right>=6\times10^{-10}M_{\odot}\,yr^{-1}$, I obtain a
\textit{maximum} dipole magnetic field at the magnetic poles of
$B_{max}\simeq1.5\times10^{10}\rm\,G$.  For $\xi=1$ the
\textit{maximum} B field becomes $B_{max}\simeq2.5\times
10^{8}\rm\,G$.  These values are within the expected range for
accreting millisecond pulsars and they match the spin down dipole
magnetic field $B_{sd}$ as well as the upper limits of
$B_{max}<3\times10^{8}\rm\,G$ proposed by \citet{tor08}.

To further check the self consistency of the accretion theory, it is useful
to obtain a \textit{minimum} magnetic field. I use the same argument
as used in \citet{psa99} and \citet{mil98}: pulsations are seen at the
outburst peak, therefore the magnetic field must be strong enough to
channel the accretion flow and enforce corotation of the accreted gas
when the accretion rate is maximum. By using the least constraining
mass accretion rate $\dot{M}=2\times10^{-9}M_{\odot}\,yr^{-1}$ (see
Section~\ref{lc}) and using Eq.~(11) in \citet{psa99} I obtain a
minimum magnetic field at the poles $B_{min}=6\times 10^{7}$ G. In
this calculation I have assumed the conservative quantities
$R=10$km, $M=1.4\,M_{\odot}$ and $\xi=1$ (here $\xi$ is what is called
$\gamma_{B}$ in \citealt{psa99}).

Both maximum and minimum magnetic fields as inferred from 
accretion theory are therefore consistent with the spin-down magnetic dipole
field obtained in Section~\ref{spin-down}.

\subsection{Motion of the Hot Spot}\label{mot}

In Section~\ref{cca} I showed that in both 2004 and in the two
weak outbursts of 2008 the coefficients of the linear correlation are
consistent with being the same, which suggests some ordered
process acting during each outbursts that provides an identical
response of the phase to a fixed perturbation of the X-ray flux.

\citet{lam08} suggested a moving hot spot on the neutron star surface
as the origin of timing noise in AMXPs. Moving hot spots were also
observed in MHD simulations of \citet{rom03} and \citet{rom04}
although the number of simulated neutron star rotations is still too
small to reach a firm conclusion.

Applying the correlation coherent technique, I find that the pulse
phases drift by approximately 0.2 cycles when the X-ray flux varies by
a factor $\sim10$ from the maximum to the minimum in 2004, and by 0.1
cycles in the two 2008 outbursts (see Figure~\ref{correlations}).
Such variations might be induced by a motion of the hot spot.  A
movement along the latitude of the neutron star will produce only
minimal changes in phase, so a drift in longitude is also required.

Under the assumption that the accretion rate tracks the X-ray
luminosity, the magnetospheric radius $r_{m}$ will move away from the
neutron star surface and approach the corotation radius $r_{co}$ as
the source luminosity drops from the peak of the outburst down to
quiescence. During this process, the hot spot can move about the
magnetic pole depending on the colatitude of the magnetic axis, as it
was observed in MHD simulations by \citet{rom03}.

In a recent work, \citet{bac10} performed detailed MHD
simulations of accreting neutron stars, and found that the hot spot
does indeed move about the magnetic pole, with small colatitudes
favoring a more pronounced motion. Although these authors could
simulate only a few rotational cycles of the neutron star, due to
computational power limitations, their results clearly show that 
the accretion flow in the pulsar magnetosphere is a highly dynamical 
process and questions the traditional picture of a static hot spot.

A detailed discussion on the modeling of the hot spot movement is
beyond the scope of this paper. A model taking into account
variations of the X-ray flux and the hot spot motion will be presented
in an accompanying paper.

\section{Spin Equilibrium and Spin Up Timescale}\label{equ}

A self-consistent measurement of the spin frequency and its long term
evolution has been given, and by knowing the approximate duty cycle
for the outbursts/quiescence cycle of IGR J00291, it is possible to
determine a timescale for the spin up of the pulsar.

I call $\Delta^{-1}$ the inverse of the duty cycle for the
outbursts/quiescence episodes, whose value is approximately $1\%$
(2004 outburst length$\sim 13$ d, 2004-2008 quiescence length$\sim
1350$d) which is a term necessary because the accretion torques act
only during the outbursts, while they are not effective during
quiescence. In this first approximate calculation I do not include the
effect of the spin down due to magnetic dipole torques in order to
obtain the shortest timescale possible. I use
$\nu_{s}\simeq600\rm\,Hz$ and
$\dot{\nu}_{su}\simeq5\times10^{-13}\rm\,Hz\,s^{-1}$, with a duty cycle
$\Delta\simeq\,0.01$.  Therefore the spin up timescale is
\begin{equation}
t_{spin-up}\equiv\frac{\nu_{s}}{\dot{\nu}_{su}}\Delta^{-1}\equiv\,4\,Gyr
\end{equation}
 which means that the pulsar will not change its spin frequency
significantly for a long timescale, or in other words that it is close
to the spin equilibrium (see for example \citealt{cam98}).

Furthermore, there is a further slow down on this timescale deriving
from the spin down observed in quiescence. The timescale for the
spin-down
($\dot{\nu}_{sd}\simeq-3\times10^{-15}\rm\,Hz\,s^{-1}$) is:
\begin{equation}
t_{spin-down}\equiv\frac{\nu_{s}}{|\dot{\nu}_{sd}|}(1-\Delta)^{-1}\simeq 8\rm\,Gyr
\end{equation}

which is of the same order of magnitude as the spin-up timescale. 
By combining these two timescales, under the hypothesis that they are both
representative of the long term behavior of the pulsar, 
the long-term timescale for the spin evolution is:
\begin{equation}
t_{spin}\equiv\left(\frac{\nu_{s}}{\dot{\nu}_{sd}\times(1-\Delta)+\dot{\nu}_{su}\times\Delta}\right)\simeq\,7\rm\,Gyr
\end{equation}
This timescale strengthens the suggestion that the pulsar in IGR
J00291 is indeed close to spin equilibrium.

\subsection{The lack of Sub-Millisecond Pulsars}\label{subms}

A very important problem in neutron star physics is how to justify the
absence of pulsars with spin period in the sub-millisecond rage.
Although only a rather limited number of sources were known at the
time, \citet{cha03} and \citet{cha05} studied the spin distribution of
AMXPs and nuclear powered pulsars and discovered that this was
consistent with a flat distribution truncated at approximately 730 Hz
with a 95\% confidence level. Although strong observational biases
exist for the detection of a radio pulsar above this frequency, X-ray
observations taken with observatories like RXTE/PCA should not be
affected by a significant loss of sensitivity at least up to 2 kHz
(see \citealt{cha08b} and references therein).  Hence it remains
unexplained why the maximum spin frequency known for accreting neutron
stars is 619 Hz~\citep{har03}.  \citet{hes06} discovered the fastest
known radio millisecond pulsar, that spins at 716 Hz, surprisingly
close to the cut-off spin limit of 730 Hz.  If I repeat the
calculation of the spin distribution cutoff as done by \citet{cha03},
with a sample size which has doubled in the meanwhile (from 11 to 23
known accreting neutron star spins, see Table~\ref{tabspin}) \textit{I
still find a cutoff of 730 Hz, but with a higher confidence level of
99\%}.

\begin{figure}[t]
  \begin{center}
    \rotatebox{-90}{\includegraphics[width=0.65\columnwidth]{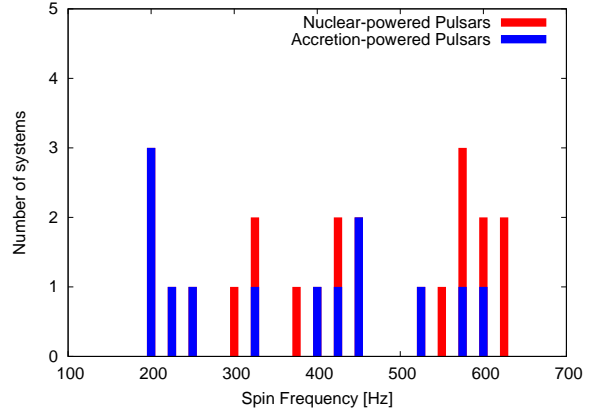}}
  \end{center}
  \caption{Spin frequency distribution. The red bars represent 10
nuclear powered pulsars whose spin frequency is measured via burst
oscillations during thermonuclear bursts. The 13 accretion powered
pulsars are identified by blue bars, and no accreting pulsar with
spin below 100 Hz is counted in the sample, according to the 
definition of millisecond pulsar. Note that the red and the
blue histograms are not overlapped, but they are added. If a pulsar is
both accretion powered and nuclear powered, I have counted it as an
accretion powered.  The total number of neutron stars in the sample
(accretion plus nuclear powered) is 23.
    \label{histogram}}
\end{figure}

IGR J00291 is the third fastest known accreting neutron star and among
the fastest neutron stars known. It is therefore at the upper end of
the spin distribution of accreting neutron stars, which is plotted in
Figure~\ref{histogram} (see \citealt{wat08},\citealt{mar07} and
\citealt{gal10} for references). If the spin evolution of this source
is representative of the behavior of AMXPs, then it explains the
existence of a cutoff of 730 Hz in the spin distribution of accreting
pulsars.

\begin{table}
  \caption{Accretion and Nuclear Powered Millisecond Pulsars }
\centering
\begin{tabular}{ccc}
\hline
\hline
Source Name & Spin Frequency [Hz]\\
\hline
Swift J1756-2508 & 182\\
XTE J0929-314 & 185\\
XTE J1807-294 & 190\\
NGC 6440 & 205\\
IGR J17511 & 245\\
IGR J17191-2821 & 294\\
MXB 1730-335 & 306\\
XTE J1814-338 & 314\\
4U 1728-34 & 363\\
HETE J1900.1-2455 & 377\\
SAX J1808.4-3658 & 401\\
4U 0614+09 & 415\\
XTE J1751-305 & 435\\
SAX J1748.9-2021 & 442\\
SAX J1749.4-2807 & 518\\
KS 1731-260 & 526\\
Aql X-1 & 550\\
EXO 0748-676 & 552\\
MXB 1659-298 & 556\\
4U 1636-536 & 581\\
IGR J00291-5934 & 599\\
SAX J1750.8-2900 & 601\\
4U 1608-52 & 620\\
\hline
\end{tabular}\\
%All errors are at $\Delta\chi^2=1$. \\
\label{tabspin}
\end{table}

Several mechanisms have been proposed to explain the lack of
sub-millisecond pulsars, but what appears certainly true is that
these spin frequencies are not limited by the break-up frequency
of neutron stars, which sets in when the centrifugal force exceeds
the gravitational pull. 
The break-up frequency depends on the equation of state of ultra-dense
matter, and its determination is therefore of fundamental importance
to understand the ground state of matter~\citep{web05}. 
Basically all equation of states allow a much higher break-up
frequency than 730 Hz. Alternative models to explain the lack of
sub-millisecond accreting neutron stars have been proposed,
including loss of angular momentum due to emission of gravitational
radiation and the existence of magnetic spin equilibrium.

In the former case, a train of gravitational waves is emitted as soon
as the neutron star develops a significant quadrupole moment.  The
angular momentum brought away by the gravitational waves slows down
the pulsar rotational period, thus preventing accreting neutron stars
to reach sub-millisecond periods. The gravitational wave torque is
proportional to $\nu_{s}^{5}$, so its effect greatly increases at high
rotational frequencies. This can produce the sharp cutoff in the spin
distribution (see~\citealt{cha08b} for further discussions of the
problem).  A loss of angular momentum via gravitational radiation is
still an open possibility, but the good agreement between the magnetic
field determination via accretion torques and via magneto dipole spin
down seems to suggest a simpler (and in this sense more likely)
explanation for the existence of a 730 Hz cutoff. 

If the timescale for the spin-up in IGR J00291 is of
the order of $7$ Gyr (see Section~\ref{sat}), then a plausible
explanation for the 730 Hz cutoff
is that accreting neutron stars reach the spin equilibrium
earlier than it is required to reach sub-millisecond periods.
In this sense, the reason why there are no observed sub-ms pulsars 
might be a simple consequence of binary and magnetic field evolution
(see also \citealt{lam05} and \citealt{lam08b}). 

It is possible to compare this behavior with the only other AMXPs with
a measured spin down in quiescence: SAX J1808.4-3658 (\citealt{har08},
\citealt{har09}, \citealt{pat09a}). The overall long term spin
frequency in SAX J1808.4-3658 is \textit{decreasing} over an observed
baseline of $\sim\,10$ years, suggesting that no significant accretion
torque operates during the outbursts. Therefore also this pulsar will
not significantly move in the spin distribution diagram on a timescale
comparable with the Hubble time. Unless the neutron star was born with
a spin already in the millisecond range, strong accretion torques must
have spun up the pulsar in the past. Its current magnetic field and
average mass accretion rate might be instead too small to allow a
significant spin up during the outbursts. Therefore the spin evolution
of these two AMXPs is compatible with a scenario in which
AMXPs evolve close to the spin equilibrium on a timescale shorter than
it is required to spin up to the sub-millisecond range. 

Before drawing a firm conclusion it is important to investigate the
spin evolution of more accreting neutron stars, but it seems justified
here to propose that, given the observed spin evolution of IGR J00291
and SAX J1808, sub-millisecond pulsars might be, at best, extremely
rare.

\section{Summary in the framework of the Recycling Scenario}\label{recy}

A first achievement for the recycling theory arrived with the
discovery of the first Accreting Millisecond X-ray Pulsar in 1998
\citep{wij98}. Another fundamental step has been the recent detection
of a millisecond radio pulsar in a position coincident with a
a previously known quiescent neutron star X-ray binary
(\citealt{arc09}, \citealt{hom06}). This was a further evidence that accreting millisecond pulsars
might indeed turn on as millisecond radio pulsars. 

However, there was still a missing test that needed to be performed
before the \textit{recycling scenario} could be accepted as the correct
theory of accreting neutron stars: the accreting pulsar is spun up, so
it must be possible to observe accretion torques in the process of
accelerating the neutron star rotation and spin down during quiescence
due to magneto dipole torques.  Many claims have been made
for a detection of an accretion torque in accreting millisecond
pulsars, but none of these has been broadly accepted until now because
of the presence of timing noise that affects the determination of spin
and accretion torques when using a standard coherent timing analysis.

The results presented here show that accretion torques are present in
IGR J00291, and the spin evolution over 4 years is entirely consistent
with the prediction of the \textit{recycling scenario}. The gas is
channeled along the weak magnetic field lines very close to the
neutron star surface, at a distance of less than 24 km from the
neutron star center. Furthermore, a slow spin down is detected when
the accretion halts (or is strongly reduced), which I ascribe to
magnetic dipole spin down as observed in radio pulsars.  This allows
the measurement of the magnetic field of the neutron star, which I
determine to be
$1.5\times10^{8}$G$\simless\,B\simless\,2\times10^{8}\pm0.3$ G, with
an uncertainty on the two extreme values of $0.3\times10^{8}$G.  This
value is consistent with that inferred from the accretion torques
during the 2004 outburst. Given the large uncertainties in the
  analysis discussed in Section 5, it is still premature to state that
  the results reported in this paper finally confirm the recycling
  scenario. However, it has been shown here that there is no need for
  new physics to explain the results reported. For example, there is
  no evidence for a spin down mechanism other than magneto dipole
  torques.

There is instead strong evidence for an ordered process that is always
present in all observed outbursts that might be ascribed to a motion
of the hot spot on the neutron star surface. Finally, I find evidence
for IGR J00291 being very close to the spin equilibrium, with the
pulsar spin evolving on timescales of $\sim7$ Gyr.  These findings
open new possibilities on the interpretation of a lack of detected
sub-millisecond pulsars: intrinsic evolutionary processes that modify
the mass transfer rate and the magnetic field might bring the pulsar
close to the spin equilibrium at a much earlier stage than it is
required to reach sub-millisecond periods.

\acknowledgements{Acknowledgments: I would like to thank Yuri Cavecchi
 for providing useful comments that helped to improve the manuscript.
 I thank Rudy Wijnands, Anna Watts, Maciej Serylak and Michiel van der
 Klis for stimulating discussions.  I am indebted to Craig Markwardt
 for pointing out the necessary corrections needed in the data
 reduction process.  I acknowledge support from the Netherlands
 Organization for Scientific Research (NWO) Veni Fellowship.}

%\bibliographystyle{aa}
%\bibliography{biblio}{99}

\end{document}